\documentclass[11pt]{article}

\usepackage{textcomp}  
\usepackage{siunitx}  
\usepackage[version=4]{mhchem}  
\usepackage{amssymb}
\usepackage{xr}  
\usepackage{caption}
\usepackage{lmodern, anyfontsize}
\usepackage[utf8]{inputenc}
\DeclareUnicodeCharacter{02BC}{\'}
\usepackage{amsmath,amsthm,bm,mathrsfs,mathtools,hyperref,lipsum}
\usepackage[numbered]{bookmark}
\usepackage{mdframed, comment}
\usepackage{courier, lineno}
\usepackage{textgreek} 
\usepackage{soul}

\usepackage{authblk}
\hypersetup{
    colorlinks,
    citecolor=black,
    filecolor=black,
    linkcolor=black,
    urlcolor=black
}
\usepackage[a4paper, total={7in, 10in}]{geometry}
\usepackage{graphicx, float, multicol, balance}
\usepackage[version=4]{mhchem}
\usepackage[sorting=none, maxnames=10, style=nature, url=false, backend=biber]{biblatex}
\usepackage{cmbright}
\title{Prebiotic magnetite enables chirality-magnetic surface feedback}

\author{
José A. P. M. Devienne$^{1}$, 
Ziwei Liu$^{1}$,
Clancy Z. Jiang$^{1}$,
Nicholas J. Tosca$^{1}$,
Thomas Ginnis$^{1,2}$
Dimitar D. Sasselov$^{3}$,
Richard J. Harrison$^{1}$,
S. Furkan Ozturk$^{4}$\thanks{ozturk@caltech.edu}\\

\normalsize{$^{1}$Department of Earth Sciences, University of Cambridge, Cambridge CB2 3EQ, UK}\\
\normalsize{$^{2}$Department of Materials Science and Metallurgy, University of Cambridge, Cambridge CB3 0FS, UK}\\
\normalsize{$^{3}$Department of Astronomy, Harvard University, Cambridge, MA 02138, USA}\\
\normalsize{$^{4}$Division of Geological and Planetary Sciences, California Institute of Technology, Pasadena, CA 91125, USA}}

\date{ }

\begin{document}
\maketitle

\begin{abstract}
    The emergence of biomolecular homochirality requires both an initial symmetry-breaking event and a mechanism to amplify and preserve a chiral imbalance. Magnetic minerals have been shown to function as chiral agents through the chiral-induced spin selectivity (CISS) effect and may have enabled homochirality on early Earth, yet the magnetic properties of magnetite formed under realistic prebiotic conditions remain unexplored. Here we show that magnetite synthesized through two geochemically plausible pathways—UV-driven photo-oxidation and nitrite-mediated oxidation of Fe(II)—produces particles dominated by single-vortex and multi-vortex magnetic domain states. Magnetic measurements and electron microscopy confirm that these populations differ markedly from the nano-fabricated thin-film substrates conventionally used in previous CISS experiments. Using 3D micromagnetic simulations, we demonstrate that single-domain and vortex-state grains undergo irreversible, exchange-driven re-magnetization when interacting with spin-polarized homochiral compounds. This magnetic irreversibility provides a robust mechanism for storing and reinforcing weak chiral bias, suggesting that prebiotic magnetite could have contributed to the emergence and stabilization of persistent chiral bias on the early Earth.
\end{abstract}

\section{Introduction}


The origin of life’s homochirality remains one of the central challenges in understanding life’s emergence. The uniform handedness of biomolecules underlies the structure and function of biopolymers such as RNA, proteins, and phospholipids, and is therefore considered an essential signature of life \cite{blackmond2019origin, sallembien2022possible, ozturk2022homochirality}. To enable the high-yielding synthesis of functional polymers, homochirality must have been established before the onset of biological replication, through a sequence involving an initial symmetry-breaking mechanism, followed by the amplification and propagation of an induced enantiomeric bias, ultimately yielding a self-sustaining homochiral network \cite{ozturk2023central, ozturk2024new, Ozturk2025Homochirality}. Understanding how life’s homochirality emerged therefore requires identifying the natural planetary environments on early Earth that could have provided the conditions necessary to achieve such an outcome \cite{sasselov2020origin}.

Various mechanisms have been proposed as potential pathways for the origin of homochirality, including parity-violating energy differences, circularly polarized light, asymmetric autocatalysis, and polarized particles produced by cosmic-ray showers, to name a few \cite{kondepudi1985weak, bailey1998circular, viedma2005chiral, rosenberg2008chiral, noorduin2009complete, globus2020chiral, meinert2011photochirogenesis, blackmond2019autocatalytic, sallembien2022possible}. Yet these processes are generally considered insufficient to account for life’s homochirality, as they generate only small enantiomeric excesses and often operate under conditions unlikely to have been present on early Earth. Moreover, they lack both an effective mechanism to amplify an initial enantiomeric imbalance and a plausible chemical pathway to transmit homochirality through stereoselective interactions, thereby enabling the emergence of a homochiral network spanning multiple classes of chiral biomolecules \cite{Ozturk2025Homochirality}. Accordingly, a plausible scenario must couple a robust mechanism for achieving network-level homochirality with processes that could occur within the material and environmental constraints of the early Earth \cite{sasselov2020origin}. 

Natural magnetic minerals—particularly magnetite ($\mathrm{Fe}_3\mathrm{O}_4$)—have recently emerged as promising candidates for facilitating chiral symmetry breaking and amplification en route to homochirality \cite{ozturk2022homochirality, ozturk2024new}. Magnetite surfaces have been shown to act as strong chiral agents through the chiral-induced spin selectivity (CISS) effect \cite{ozturk2023crystallization}, which couples electron spin polarization to molecular chirality, thereby enabling asymmetric processes controlled by electron spin \cite{ben2017magnetization, banerjee2018separation, Koplovitz2019, bloom2024chiral}. Recent experiments demonstrate that magnetized magnetite surfaces can serve as chiral nucleation templates, promoting the enantioselective crystallization of a central RNA precursor, ribose-aminooxazoline (RAO), and driving the system to full homochirality from initially racemic starting materials \cite{ozturk2023crystallization}. Once an enantiomeric bias is established, enantiopure RAO crystals can, in turn, remagnetize the underlying magnetic surface, reinforcing magnetic asymmetry and enabling a feedback mechanism between surface magnetization and enantiomeric enrichment \cite{ozturk2023magnetization}. Moreover, homochirality at the level of \textit{D}-RAO can propagate to \textit{L}-peptides and subsequently to key metabolites, supporting the emergence of a homochiral prebiotic reaction network \cite{wu2021interstrand, ozturk2023central, kim2025stereoselectivity, Ozturk2025Homochirality}.

\begin{figure}[h!]
    \centering
    \includegraphics[height=0.6\textwidth, width=\textwidth]{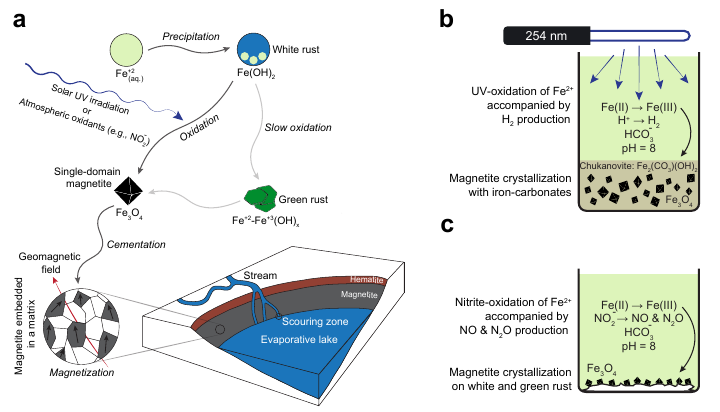}
    \caption{Magnetite formation under prebiotically plausible conditions. \textbf{a} Envisioned evaporative lake setting on early Earth, where magnetite likely formed via authigenic precipitation facilitated either by solar irradiation or nitrogen oxide from the decomposition of abiotic nitrogen fixation products \cite{buessecker2022nitrite}. As these magnetite particles form, they acquire a natural remanent magnetization under the geomagnetic field. \textbf{b} Experimental setup used to reproduce early Earth conditions and investigate magnetite formation under UV-facilitated synthesis (samples UV1 to UV3) and \textbf{c} via nitrite oxidation of Fe(II) (samples NUV8 and NUV9). Samples NUV5 to NUV7 were synthesized by mixing Fe(II) and Fe(III) with different proportions (10\%/90\%, 50\%/50\% and 90\%/10\% of Fe(II)/Fe(III), respectively).}
    \label{fig:figure1_lake_and_experiments}
\end{figure}
\begin{figure}[h!]
    \centering
    \includegraphics[width=\textwidth]{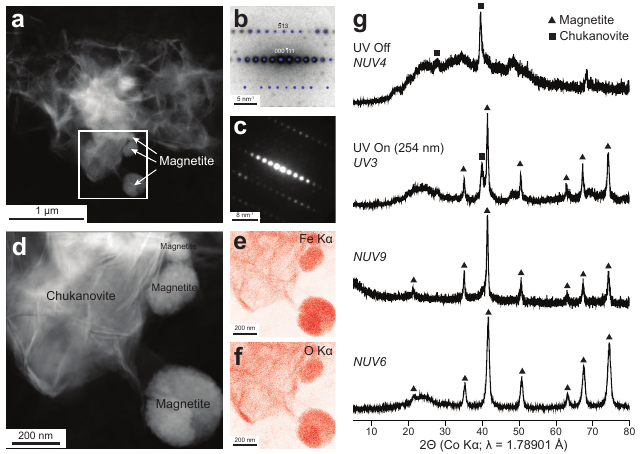}
    \caption{Different prebiotic synthesis routes produce magnetite within a range of sizes and morphologies. \textbf{a} Transmission electron microscopy imaging from a UV-synthesized sample (UV2) showing nearly equidimensional, approximately 250 nm size magnetite particles embedded in acicular chukanovite (Chk) and goethite (Go) structures. \textbf{b} and \textbf{c} are electron diffraction data confirming magnetite structure and the detected crystal planes. \textbf{d} Zoomed-in region (white square in \textbf{a}) from where energy dispersive spectroscopy (EDS) chemical maps were taken, highlighting regions rich in \textbf{e} iron  and \textbf{f} oxygen. \textbf{g} X-ray diffraction illustrating the impact of absence (UV Off) and presence of Fe(II) oxidants (UV3, NUV9 and NUV6). Absence of oxidants completely prevents magnetite to form (NUV4); non-UV synthesized samples (NUV9 and NUV6) show clear magnetite peaks; a single evident chukanovite peak is also observed in both samples, while UV synthesis yields much clearer peaks for this phase (in addition to other iron oxide phases).}
    \label{fig:figure2_tem_eds_xray}
\end{figure}

While these findings demonstrate that magnetite can facilitate enantioselective processes, a crucial next step is to elucidate plausible formation pathways of magnetite under the prebiotic conditions of early Earth and to determine its magnetic properties in relation to spin-selective processes. Magnetite is known to form in various Fe(II)-rich environments and has been proposed as a primary precipitate from the Archean ocean \cite{li2017formation, usman2018magnetite, toscaetal2018, tabata2021experimental}. However, it remains important to investigate how environmental oxidants and solar UV irradiation contribute to its authigenic formation under the lacustrine conditions of early Earth’s surface (Figure \ref{fig:figure1_lake_and_experiments}), where selective prebiotic chemistry can generate chiral molecules with high yields, including RAO \cite{patel2015common, xu2020selective}. Previous work \cite{ranjan2019nitrogen} suggests that the concentration of nitrite on early Earth might have exceeded 1~\textmu M in the oceans and could potentially have been much higher (near-millimolar) in ponds. Furthermore, to accurately simulate aqueous surface environments on early Earth, dissolved inorganic carbon (DIC) needs to be considered as well. While estimates for early Earth atmospheric \ce{CO_2} vary by several orders of magnitude \cite{Catling2020}, and post-impact scenarios are considered to transiently produce conditions with extremely low DIC \cite{zahnle2020creation, wogan2023origin}, the presence of dissolved carbonate is known to strongly influence iron geochemistry by partitioning dissolved iron into iron-carbonate minerals. In this work, we study two plausible prebiotic magnetite formation mechanisms: the photo-oxidation of Fe(II)-carbonates and their chemical oxidation by nitrite (Figure \ref{fig:figure2_tem_eds_xray}).


Moreover, a key objective is to understand how distinct formation pathways influence the morphology and magnetic properties of the resulting magnetite, and how these properties, in turn, affect spin-selective interactions between chiral molecules and magnetite. It is well known that a magnetic material’s response to external stimuli—applied fields, temperature changes, or remagnetizing agents such as chiral molecules—is fundamentally controlled by the domain state of its magnetic carriers \cite{dunlop2001rock}. The domain state reflects the organization of magnetic moments within individual particles and depends primarily on grain size and shape. In the smallest particles, exchange interactions dominate the magnetic free energy, enforcing uniform moment alignment characteristic of a single-domain (SD) state (Figure \ref{fig:fig3_mag_props}a, left). As particle size increases, this uniform configuration becomes energetically unfavorable, giving rise to non-uniform spin textures such as single- (SV) or multi-vortex (MV) states, where magnetization curls around one or more vortex cores to reduce demagnetizing energy (Figure \ref{fig:fig3_mag_props}a, center). With further growth, vortex structures give way to domain walls separating oppositely magnetized regions, defining the multi-domain (MD) state (Figure \ref{fig:fig3_mag_props}a, right). These configurations exhibit distinct magnetic properties—including coercivity, remanence, and magnetic stability—and thus behave variably under external magnetic influence. Understanding the distribution of domain states in prebiotically synthesized magnetite, as provided in our present manuscript, is therefore essential for evaluating whether chirality-induced spin interactions observed in laboratory settings could plausibly occur under early Earth conditions.

Connecting magnetic domain-level properties to molecular-scale, spin-selective chiral interactions requires an additional analysis. Magnetic domain properties of magnetite determine not only its bulk response to external fields but also the local spin environment at its surface. It is precisely at this interface that chiral molecules would interact with magnetization. Thus, any attempt to assess magnetite’s potential as an enantioselective or symmetry-breaking agent must link macroscopic domain structures to the microscopic physics governing molecule–surface spin coupling (Figure \ref{fig:figure4_micromag_results}). Recent computational work on magnetic surfaces interacting with chiral molecules has largely operated at atomistic scales with ab initio modeling, where the quantum-mechanical nature of CISS-driven spin interactions can be explicitly treated \cite{safari2024enantioselective}. These models are valuable for estimating interaction strengths between chiral molecules and magnetic surfaces, but they remain restricted to small spin ensembles due to computational cost. As a result, they cannot assess how localized interactions between magnetic grains and a few spin-polarized chiral molecules might influence the magnetic domain state of an entire particle—a structure that emerges from the collective behavior of a macroscopic number of spins. However, determining whether and how chiral molecules can influence the magnetic structure of magnetite grains of realistic length and time scales is essential for evaluating magnetite’s potential role as a chiral symmetry-breaking agent on prebiotic Earth. To address this, here we investigate the interaction of chiral molecules with real-sized prebiotic magnetite grains that we produce in our experiments, using micromagnetic modeling to evaluate the feasibility of spin exchange–driven re-magnetization by chiral molecules (Figure \ref{fig:figure4_micromag_results}). Even for conservative estimates of the spin-exchange interaction, a significant portion of magnetite grains can undergo irreversible magnetization by chiral molecules (Figure \ref{fig:figure4_micromag_results}e), and retain their remagnetized state over geological time scales \cite{nagy2017stability}. This process enables a persistent physical chiral bias of hemispheric-scale uniformity with a long lifetime for prebiotic chemistry in the geochemical environment, despite thermal fluctuations and geomagnetic reversals.

\section{Results}

\subsection{Magnetite synthesis under prebiotically plausible conditions}

Magnetite was synthesized through four separate pathways. In all four cases, the experimental conditions (DIC = 20 mmol kg\textsuperscript{-1}) correspond to an equilibrium of approximately 10\textsuperscript{-2.3} bar of carbon dioxide in the atomsphere. This value is within the range of plausible estimates of the level of the \textit{p}CO\textsubscript{2} for the Hadean Earth \cite{Catling2020, zahnle2020creation, wogan2023origin}. In the first approach, FeCl\textsubscript{2} and FeCl\textsubscript{3} solutions were mixed in a 400 mL working matrix prepared (see Methods) in a 500 mL \textit{Pyrex} beaker. After homogenization, the pH was adjusted to 8.0 with 1 mol kg\textsuperscript{-1} NaOH added drop wise. The mixture was stirred for 4 hours, transferred to a \textit{Duran} bottle, sealed with a polypropylene cap, and incubated for one week. The resulting precipitate was collected by filtration through 0.22 \textmu m membranes and dried in a low-humidity glove box for 24 hours. Although effective, this method is unlikely to represent prebiotic conditions, as a rapid pH increase from 3 to 8 and the availability of FeCl\textsubscript{3} are both unrealistic for early Earth environments.

To better approximate prebiotic scenarios, a second method involved adding FeCl\textsubscript{3} dropwise into FeCl\textsubscript{2} while maintaining pH 8 \textit{via} simultaneous NaOH titration. This procedure mimics the gradual formation of Fe(III) through slow oxidative processes. On early Earth, plausible oxidants include UV irradiation and nitrite \cite{ranjan2019nitrogen, green2021illuminating}. Therefore, in the third and fourth methods, Fe(II) solutions were partially oxidized either by UV exposure or by treatment with dilute sodium nitrite. Both oxidation pathways successfully produced magnetite (Figure \ref{fig:figure2_tem_eds_xray}). The successful formation of magnetite under these relatively high dissolved CO\textsubscript{2} conditions—where competitive partitioning into carbonates is significant—further strengthens the prebiotic plausibility of these synthesis pathways.

\subsection{Magnetic properties of prebiotic magnetite}

\begin{figure}[h!]
    \centering
    \includegraphics[width=0.89\textwidth]{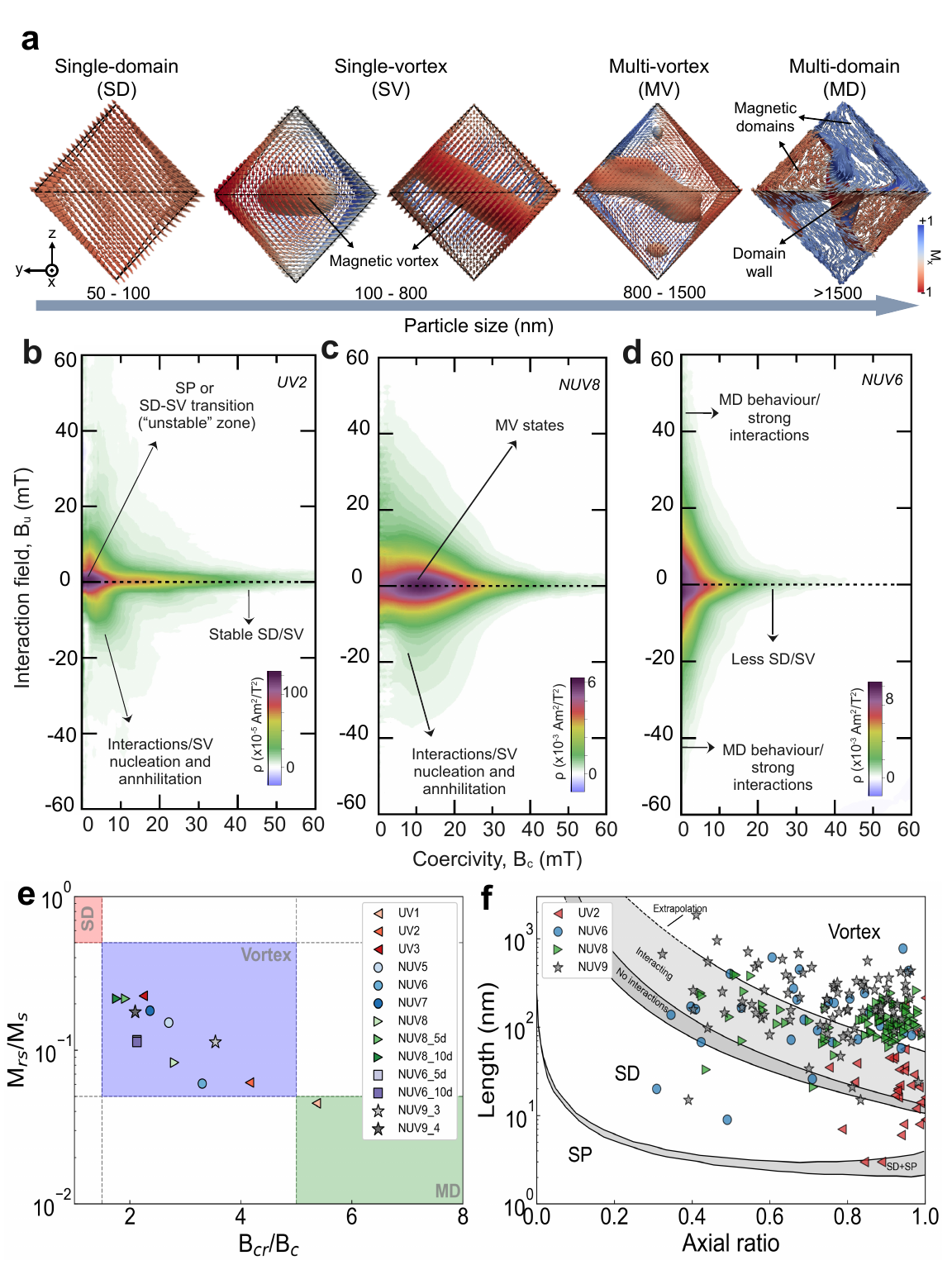}
    \caption{Magnetic domain states in prebiotically plausible magnetite. \textbf{a} Different domain states obtained with micromagnetic modeling of individual magnetite particles with different sizes: single-domain (SD), single-vortex (SV), multi-vortex (MV) and multi-domain (MD) states. \textbf{b} First-order reversal curves measurements for samples UV2 (UV-driven synthesis), \textbf{c} NUV8 (NO\textsubscript{2}\textsuperscript{-}-driven synthesis at pH 8.0) and \textbf{d} NUV6 (Fe(II) and Fe(III) mixture without pH controlling). \textbf{e} Day plot \cite{williams2024dayplot} of the ratio between saturation remanent ($M_{rs}$) and saturation ($M_s$) magnetizations against the ratio between remanence coercivity ($B_{cr}$) and coercivity ($B_c$); light red, blue and green regions are those where SD, SV and MD particles fall within, respectively \cite{Day1977Hysteresis, williams2024dayplot}. \textbf{f} Butler-Banerjee plot delineating a two-dimensional size-shape range for SD and non-SD (i.e., SV, MV and MD) particles of magnetite \cite{butler1975}.}
    \label{fig:fig3_mag_props}
\end{figure}

Bulk magnetic characterization measurements, including hysteresis, backfield curves, temperature-dependence of magnetic susceptibility and first-order reversal curves (FORC), indicate that prebiotic synthesis of magnetite produce particles covering virtually the whole spectrum of domain states (Figure \ref{fig:fig3_mag_props}). In the Day plot \cite{Day1977Hysteresis}, most samples cluster within the vortex region, suggesting that non-uniform magnetic structures dominate the bulk magnetic signal (Figure \ref{fig:fig3_mag_props}e). Nevertheless, the Day plot alone cannot resolve the coexistence of subpopulations with distinct domain structures, and thus FORC measurements were helpful to provide more detailed insights into their magnetic behavior. In UV‐synthesized samples, FORC diagrams (Figure \ref{fig:fig3_mag_props}b) show a pronounced central ridge along the coercivity axis ($B_c$) extending up to 60 mT, diagnostic of stable single‐domain ($\sim$50–100 nm) and single‐vortex ($\gtrsim$100 nm) grains \cite{roberts2014understanding, nagy2024micromagnetic}. The nearly symmetrical spread along the interaction‐field axis ($|B_u| \lesssim 10$ mT) indicates weak to moderate magnetostatic coupling, while diagonal features reflect the nucleation and annihilation of vortex states on application of external fields \cite{roberts2014understanding, nagy2024micromagnetic}. UV-synthesized samples also display a noticeable low‐coercivity component near $B_c\approx0$, pointing to the coexistence of thermally unstable, superparamagnetic (SP; $<$50 nm) grains and “critical‐size” particles at the SD–SV threshold, where flattened magnetic energy landscapes produce very low stability \cite{nagy2017stability, ge2021models}. Contrastingly, samples produced via nitrite-oxidation of ferrous iron display a large, central coercivity peak around $\sim$13.7 mT (Figure \ref{fig:fig3_mag_props}c), characteristic of particles adopting multi-vortex states \cite{lascu2018vortex}, while samples produced from mixtures of Fe(II) and Fe(III) display near-zero coercivity peaks and signals spreading considerably along the interaction field axis, likely as an indicative of strong interactions \cite{Harrison2014FORCulator} and/or the presence of multi-domain particles \cite{roberts2017resolving} (Figure \ref{fig:fig3_mag_props}c).

Electron microscopy and energy-dispersive spectroscopy analyses further support the FORC interpretations by providing detailed morphological and compositional constraints on the synthetic magnetite. UV-driven synthesis produced a complex architecture of nearly equidimensional magnetite grains, $\sim$15–200 nm in size (within the SP-SD–SV range; Figure \ref{fig:uv2_size_dist}), embedded within an acicular matrix containing chukanovite and goethite (Figure \ref{fig:figure2_tem_eds_xray}b–d, f, \ref{fig:xrd_si}). Notably, these particle sizes falls precisely within the range where single-vortex particles are expected to form according to high-resolution 3D micromagnetic modeling \cite{nagy2017stability, nagy2019nano, nagy2024micromagnetic}, further supporting the interpretation of FORC measurements for UV-synthesized samples. On the other hand, imaging of non-UV-synthesized samples revealed larger aggregates, with individual magnetite particles reaching up to $\sim$1.0 $\mu$m in length, consistent with the MV–MD domain states inferred from the magnetic data (Figure \ref{fig:em_uv6_si}). Projecting the reconstructed size and shape distributions onto the Butler–Banerjee plot (Figure \ref{fig:fig3_mag_props}f), which relates particle length to axial ratio and predicts domain states based on particle geometry \cite{butler1975, muxworthy2008}, provides a quantitative view of these textural variations. In our dataset, particle dimensions range from a few tens of nanometers to over one micrometer, with axial ratios between $\sim$0.35 and 1.0, spanning nearly the entire size–shape space of the diagram but mostly concentrated on the vortex regions. Most UV-synthesized grains fall within the non-interacting zone, consistent with SV-like configurations \cite{Nikolaisen2020}, while the larger and more elongated grains formed through nitrite-mediated synthesis extend into the vortex field. Together, these observations confirm that both synthesis routes produce magnetite populations encompassing a broad size/shape distribution, but are mostly dominated by particles within the vortex size range.

\subsection{Exchange-driven irreversible magnetization changes caused by chiral molecules}

\begin{figure}[h!]
    \centering
    \includegraphics[width=\linewidth]{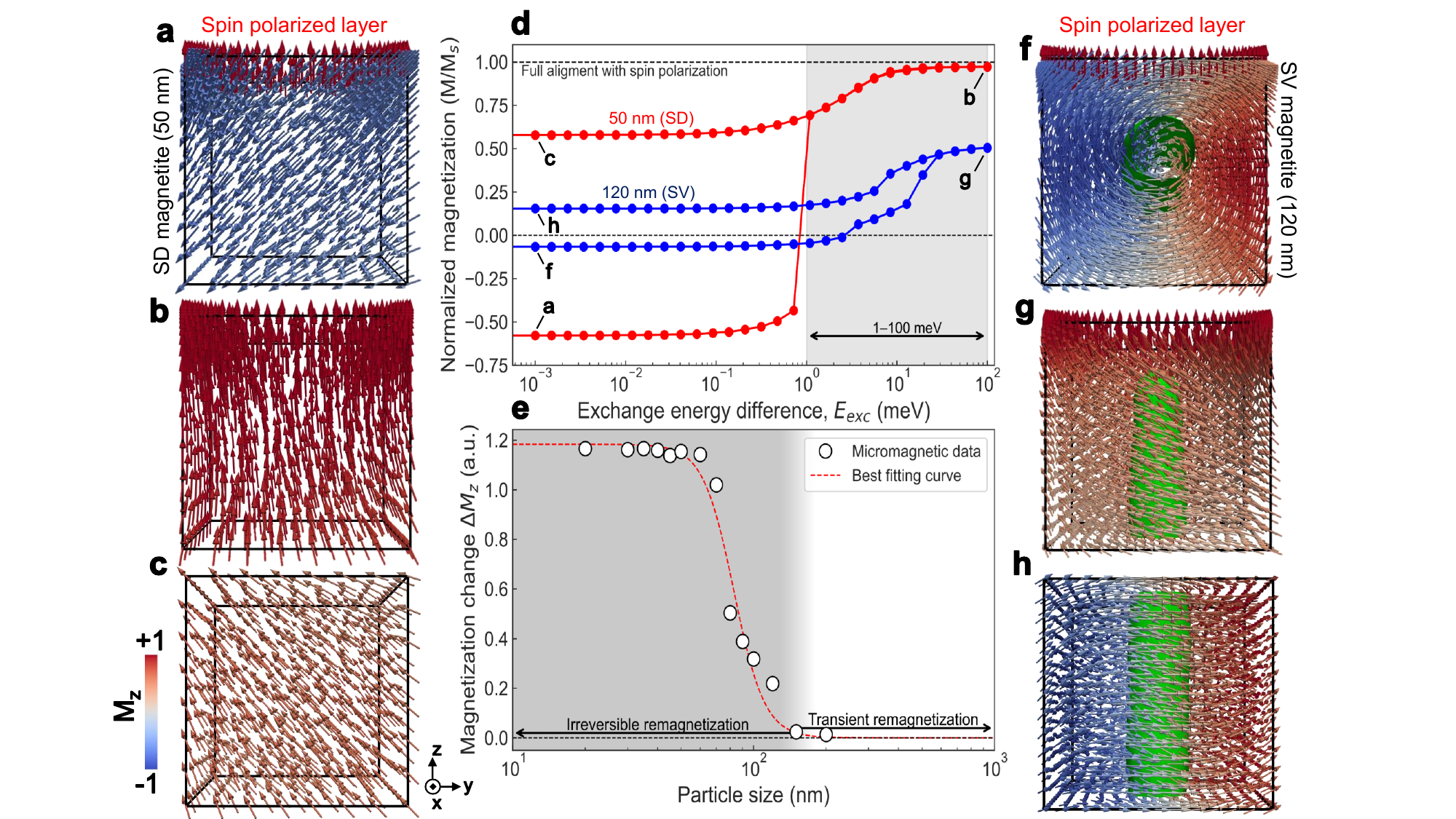}
    \caption{Chiral molecules can irreversibly remagnetize single-domain and single-vortex states in magnetite particles. \textbf{a-c} Changes in the magnetic domain structure of a 50 nm, single-domain cube of magnetite caused by a layer of spin-polarized molecules (red arrows at the top a,b and f,g) as the strength of the exchange interaction ($E_{exc}$) between the spin-polarized layer and magnetite is varied from 0 to 100 mev and back from 100 meV down to 0. \textbf{a} Initial SD state aligned with the [$\Bar{1}$,$\Bar{1}$,$\Bar{1}$] direction; \textbf{b} intermediate remagnetized SD state aligned with the [0,0,1] direction (i.e., almost fully aligned with the layer's spin polarization); \textbf{c} final SD state aligned with the [$\Bar{1}$,$\Bar{1}$,1] direction after the exchange interaction strength is set back to zero. \textbf{d} Shows how the z-component of magnetization (i.e., the one along the fixed molecular spin polarization orientation) changes as a function of exchange interaction strength for a 50 nm sized, SD magnetite particle (red curve) and a 120 nm sized, SV particle (blue curve). \textbf{e} Magnetization change (defined as the difference between the z-components of the initial and final states after cyclically varying $E_{exc}$) as a function of grain size; particles with sizes below $\sim$130 nm undergo irreversible remagnetization, while larger particles are only transiently remagnetized when exposed to the spin polarized molecules. \textbf{f-h} Changes in the magnetic domain structure for a 120 nm, single-vortex magnetite particle: \textbf{f} initial SV aligned with [1,0,0] direction; \textbf{g} intermediate SV aligned with the [0,0,1] direction with a partially destroyed vortex as $E_{exc}$ is at its maximum value, and \textbf{h} final [0,0,1] SV state with the vortex structure restored as the $E_{exc}$ is set back to zero.} 
    \label{fig:figure4_micromag_results}
\end{figure}

We employed the 3D micromagnetic modeling package MERRILL \cite{OConbhui2018MERRILL}, which incorporates quantum-mechanical exchange effects via a continuum approximation, to assess whether exchange-driven interaction with spin-polarized homochiral materials can induce any changes in magnetization structure of magnetite particles (Figure \ref{fig:figure4_micromag_results}). Our micromagnetic results indicate that both single-domain (Figure \ref{fig:figure4_micromag_results}a-d) and single-vortex magnetite particles (Figure \ref{fig:figure4_micromag_results}e-h) undergo remagnetization when exposed to spin-polarized materials. By cyclically varying the exchange interaction between spin-polarized molecules and magnetite particles from 0 to 100 meV (i.e., within the upper limits of $\approx$0.01-100 meV usually reported for such interactions \cite{ozturk2023magnetization}), our model was able to obtain exchange-driven "hysteresis" curves  that confirm that the domain state changes in SD and SV states (up to $\sim$100-130 nm) are irreversible, as indicated by the abrupt "jumps" observed in the lower branches of the "exchange-hysteresis" curves. As a consequence, once remagnetized by a spin-polarized material, SD and SV magnetite particle do not return to their initial state even after the removal of the molecules: an SD particle initially magnetized along the [$\Bar{1}$,$\Bar{1}$,$\Bar{1}$] direction (Figure \ref{fig:figure4_micromag_results}a) will end up magnetized in the [$\Bar{1}$,$\Bar{1}$,1] direction (Figure \ref{fig:figure4_micromag_results}c) when the exchange interaction energy is cyclically varied from 0-100 meV and back from 100-0 meV. Similarly, an initial SV state magnetized along the [1,0,0] direction (Figure \ref{fig:figure4_micromag_results}f) will finally align with the [0,0,1] direction (Figure \ref{fig:figure4_micromag_results}h) as the interaction energy with the molecules is varied cyclically. Notably, both domain states end up with a larger contribution of their magnetization along the molecules' spin polarization (namely, their $M_z$ component, Figure \ref{fig:figure4_micromag_results}d,e). Therefore, spin-polarized molecules remagnetize SD and SV magnetite particles in a way to maximize their alignment with the the molecules' spin polarization. This, coupled with the fact that the magnetization changes we observed with our models are irreversible, indicate that on early Earth settings, SD and SV magnetite particles are capable of providing a robust platform for developing and propagating a consistent chiral bias towards subsequent generations of homochiral crystals.

\section{Discussion}

\subsection{Relevance of magnetic domain states in chirality-induced magnetic interactions}

The potential of magnetite to act as a natural symmetry-breaking agent for chiral molecules has recently been highlighted by experiments showing that multi-domain (MD) magnetite thin films can induce enantiomeric excesses exceeding $\sim$60\% during the crystallization of racemic RAO, a key RNA precursor \cite{ozturk2023crystallization}. Furthermore, RAO itself can trigger avalanche-like remagnetization of these films, propagating magnetic changes well beyond the initial contact zone \cite{ozturk2023magnetization}. Together, these studies reveal a dual mechanism by which magnetite can (i) break chiral symmetry via its natural remanent magnetization acquired under the Earth`s magnetic field, and (ii) amplify this bias through feedback between homochiral molecules and magnetic surfaces.

Despite these advances, experimental demonstrations of chirality-induced magnetization changes have so far been confined to the ends of the magnetic size spectrum (Figure ~\ref{fig:fig3_mag_props}a): nanometer-scale superparamagnetic (SP) particles ($\lesssim$10-30 nm) \cite{Koplovitz2019, safari2022enantiospecific} and micrometer-scale MD thin films ($\gtrsim$1.5–3.0 µm) \cite{ziv2019afm, ozturk2023magnetization}. The wide intermediate regime, spanning sub-micron to micron-sized grains that adopt single-vortex (SV) and multi-vortex (MV) states, has remained unexplored, even though such grains overwhelmingly dominate terrestrial igneous and sedimentary environments \cite{roberts2017resolving, Nikolaisen2020}, as well as extraterrestrial materials \cite{Gattacceca2012martian, Cao2024lunar}.

Our results fill this critical gap: we show that under prebiotically plausible conditions, both UV-driven and nitrite-mediated magnetite synthesis pathways produce particles whose sizes and shapes fall mostly within the vortex state range. FORC analyses, electron microscopy imaging, and the reconstructed size–shape distributions (Figure \ref{fig:uv2_size_dist}-\ref{fig:nuv9_size_dist}) demonstrate that these grains form the majority of the magnetic assemblage, with SD–SV particles dominating the UV pathway and larger SV–MV grains characterizing the nitrite pathway. Thus, the geochemical processes likely responsible for magnetite formation on the early Earth produce predominantly the domain states that have remained untested in previous CISS-related studies.

\subsection{Irreversibility: The key for effectively propagating chiral bias}

In our model, we consider a conservative exchange coupling between the spin-polarized molecular layer and the magnetite surface, in which the molecule-surface interaction is parameterized as an effective interfacial exchange bias, expressed as a per-interaction energy difference ($\Delta E$) ranging from 0--100~meV (see Methods). Notably, AFM-based spin-exchange microscopy of chiral molecules on ferromagnetic films has identified that $\Delta E$ can be as large as 150~meV \cite{ziv2019afm}. Our simulations show that substantially smaller molecular exchange energies ($\Delta E \approx 0.1$--30~meV; see Figure \ref{fig:figure4_micromag_results}d) are already sufficient to bias or reconfigure single-domain and vortex magnetite particles within the 20--170~nm size range. Consequently, the experimentally observed spin-exchange energy of $\approx$150~meV per interfacial interaction is more than sufficient, in principle, to generate the magnetic bias required to drive magnetization changes within this size range.

Further support stems from observations of chirality-induced magnetization switching in metallic thin films. Adsorption of a chiral self-assembled molecular layer has been shown to produce magnetization reversal in Co/Pt multilayers with an effective energy transfer on the order of meV per electron \cite{ben2017magnetization}. Similarly, molecular layers adsorbed on Ni thin films have been shown to induce magnetization switching equivalent to an applied field of $\sim$0.4~T \cite{Dor2013ChiralMemory}, while in magnetite thin films a $\sim$~2.5~mT effective field was estimated to be caused by the adsorption of RAO molecules on it \cite{ozturk2023magnetization}. Although these experiments utilize diverse materials and length scales, they consistently indicate that CISS-induced spin exchange at molecular interfaces can supply at least meV-scale biases and, in some cases, field-equivalent perturbations of the order of mT, far exceeding the geomagnetic field (tens of $\mu$T).

The key finding obtained with our models is that both SD and SV magnetite grains undergo irreversible remagnetization when exposed to spin-polarized homochiral materials (Figure \ref{fig:figure4_micromag_results}). Critically, this means that once the exchange interaction between chiral molecules and the magnetic surface is cycled back to zero (i.e., mimicking when the crystals are removed or washed off), the magnetization does not return to its initial state. Instead, particles retain a remanence component aligned with the molecules’ spin polarization. In SD grains, this results in a switch between crystallographic easy axes; in SV grains, it induces vortex-core realignment and reorientation of the net moment. In both cases, exchange-driven “hysteresis” demonstrates that even weak spin-polarized interactions can induce irreversible changes in realistic magnetite grains. 

One should note, however, that our quasi-static, conservative micromagnetic approach does not explicitly include thermal activation. The purpose of the simulations is not to model the kinetics or time scales of magnetization switching, but rather to determine how a spin-polarized interfacial exchange bias reshapes the magnetic free-energy landscape and alters the relative stability and accessibility of competing domain states. Thermal fluctuations primarily affect the rate at which a particle can overcome energy barriers separating these states, but they do not change the existence or ordering of the minima themselves. In this sense, our results establish that the chiral exchange perturbation can drive the system across an irreversible boundary into a distinct remanent minimum, and the inclusion of thermally activated processes would then govern how rapidly such transitions occur in nature. 

A natural concern is that, once the molecules are removed (i.e., once the interfacial coupling is cycled back to zero), the remagnetized states could gradually relax back toward the original configuration over time. For the vortex-bearing grains that dominate much of our size range, available estimate of relaxation time  argue strongly against such decay on any relevant timescale: as shown in Nagy et al (2016), vortex states in magnetite grains larger than $\sim$~100~nm have relaxation times exceeding the age of the Solar System \cite{nagy2017stability}. This provides an important physical basis for interpreting the exchange-driven ``hysteresis'' observed in our simulations: once the interfacial bias has reconfigured a vortex state, the resulting remanence is expected to remain effectively locked in even after the molecular layer is removed. For single-domain grains, thermal activation can in principle produce viscous relaxation when energy barriers are sufficiently small (i.e., for particles around the SP-SD threshold, 20-30 nm). Including thermal activation would not invalidate the existence of the biased remanent minima identified here; instead, it would be expected to facilitate relaxation toward the lower-energy, molecule-selected configurations over time when barriers permit.

Importantly, the same reasoning implies that the resulting remanence should be robust to geomagnetic reversals. A reversal of the geomagnetic field changes the direction of a weak Zeeman bias (tens of $\mu$T), but this is orders of magnitude smaller than the coercive fields typical of SD–SV magnetite (mT-scale, as reflected by the hysteresis parameters in Table S1). Consequently, a reversing ambient field would be insufficient to reconfigure these remanent states unless additional processes (e.g., prolonged heating toward blocking conditions or exceptionally long-term viscous relaxation) substantially lower energy barriers. Therefore, once established, CISS-driven remanence components are expected to persist through polarity flips, allowing the magnetic asymmetry generated by homochiral molecular interactions to survive across repeated geomagnetic reversals.

This irreversibility is central to the plausibility of chiral feedback mechanisms under natural conditions. Unlike MD thin films, which require strong magnetic fields to induce a non-zero net magnetization, SD–SV grains formed in Fe(II)-rich lakes or sedimentary basins would already carry a weak but statistically coherent natural remanent magnetization acquired during growth or deposition under the geomagnetic field \cite{dunlop2001rock}. Our results show that homochiral molecules such as RAO could reinforce and amplify this inherited bias by selectively remagnetizing particles in the direction of their own spin polarization. Because the induced changes are stable at zero effective field, the magnetic asymmetry produced by these interactions would persist across dissolution–recrystallization cycles and episodic mixing. Taken together, these findings reveal a coherent, geologically plausible pathway by which prebiotic magnetite could have contributed both to the initial symmetry-breaking and subsequent amplification required to achieve homochirality.

\section{Conclusion}

Our findings demonstrate that authigenic magnetite produced through prebiotically plausible pathways provides a robust geochemical basis for the emergence and persistence of biomolecular homochirality. We show that UV-driven and nitrite-mediated oxidation produce magnetite populations dominated by single-vortex and multi-vortex grains. In addition, our 3D micromagnetic simulations demonstrate that these magnetite grains undergo irreversible, exchange-driven remagnetization when interacting with homochiral materials via the CISS effect. This irreversible exchange hysteresis provides a mechanism for reconfiguring the magnetization of magnetic minerals even for weak exchange energy differences arising from chiral molecule–magnetic surface interactions. This enables magnetite to store a persistent chiral bias relevant to biomolecular homochirality. When coupled with a geomagnetic field that induces uniform remanence across a planetary hemisphere, this process establishes a self-reinforcing feedback loop: magnetite templates enantioselective crystallization, while the resulting enantiomerically enriched materials amplify and stabilize the mineral’s magnetic bias. In summary, these results suggest that magnetite could have functioned as a persistent chiral agent enabling homochirality on prebiotic Earth.

\section*{Acknowledgments}

The authors thank the Origins Federation for funding and its members for fruitful discussions. S.F.O. and D.D.S. acknowledge the Harvard Origins of Life Initiative; S.F.O. further acknowledges support from the Kavli-Laukien fellowship program, Caltech startup funds, the William H. Hurt Scholarship Program, and the Fellows of King's College Cambridge for their hospitality. R.J.H., N.J.T. and J.A.P.M.D. acknowledge the Leverhulme Centre for Life in the Universe (LCLU) funding (LBAG/329), with additional LCLU support for N.J.T. and Z.L. (RC-2021-032). Part of this work was performed by J.A.P.M.D. as a visiting researcher at the Institute for Rock Magnetism (IRM) at the University of Minnesota, which is supported by the National Science Foundation (NSF EAR-2153786) and the University of Minnesota. T.G. acknowledges the Wolfson Electron Microscopy suite for the use of the TF Spectra 300. Finally, the authors are grateful to Dr. Rosa Danisi and Dr. Iris Buisman for assistance with electron microscope imaging, and to Dr. Elias Mansbach for insightful discussions.

\newpage
\printbibliography

\newpage

\section*{Methods}

\subsection*{General Method}

Reagents and solvents were obtained from \textit{MP Biomedicals, Sigma-Aldrich}. Reagents and solvents were used without further purification unless otherwise stated. A \textit{Metrohm} 902 Titrando autotitrator combined with a \textit{Metrohm} pH electrode was used to measure and adjust the pH to the desired value. All solutions were prepared in a \textit{Coy} polymer glovebox filled with 5\% H\textsubscript{2}/95\% N\textsubscript{2} gas and maintained under anoxic conditions by a pair of palladium catalysts and anhydrous CaSO\textsubscript{4} desiccants. All stock and working solutions were prepared using deionised water that was first deoxygenated by purging with O\textsubscript{2}-free N\textsubscript{2} gas for at least 1 h \cite{butler1994removal}, then placed on a stir plate for 24 h in the glovebox to allow the residual O\textsubscript{2} to escape and be removed by the catalysts. FeCl\textsubscript{2} and FeCl\textsubscript{3} stock solutions (0.5 mol kg\textsuperscript{-1} were prepared by dissolving analytical grade powders in the deoxygenated deionised water. 

\subsection*{Working matrix preparation} 

NaHCO\textsubscript{3} solutions were freshly prepared for each experiment to ensure that the total dissolved inorganic carbon (DIC, 20 mmol kg\textsuperscript{-1}) across experiments was not affected by the degassing of the solution over time. The stoichiometric ionic strength (\textit{I} = 0.1 mol kg\textsuperscript{-1}) of all experiments was balanced by adding predetermined amounts of NaCl.

\subsection*{Magnetite synthesis by mixing FeCl\textsubscript{2} and FeCl\textsubscript{3} solution}

The working matrix (400 mL) was prepared in a 500 mL \textit{Pyrex} glass beaker. After the solution was homogenized by stirring, predetermined FeCl\textsubscript{2} and FeCl\textsubscript{3} solution solution was added (Table \ref{sample_prep}, entry 5, 6, 7). The pH was adjusted to 8.0 by adding NaOH (1 mol kg\textsuperscript{-1}) solution drop-wise. The resulting solution was stirred for 4 hours and then transferred to a \textit{Duran} glass bottle and sealed with a polypropylene cap.  After one week of incubation, the precipitate was collected by filtration through 0.22 \textmu m membranes and dried in the low-humidity glovebox for 24 hours.

\subsection*{Magnetite synthesis by titrating FeCl\textsubscript{2} and FeCl\textsubscript{3} solution}

The working matrix (400 mL) was prepared in a 500 mL \textit{Pyrex} glass beaker and 3.6 mL FeCl\textsubscript{2} stock solution was added (Table \ref{sample_prep}, entry 8). The pH was adjusted to 8.0 by adding NaOH (1 mol kg\textsuperscript{-1}) solution drop-wise. After the solution was homogenized by stirring, 0.4 mL FeCl\textsubscript{3} stock solution and NaOH (1 mol kg\textsuperscript{-1}) solution were added into the solution drop-wise simultaneously to maintain the pH at 8.0. The resulting solution was stirred for 4 hours and then transferred to a \textit{Duran} glass bottle and sealed with a polypropylene cap.  After one week of incubation, the precipitate was collected by filtration through 0.22 \textmu m membranes and dried in the low-humidity glovebox for 24 hours.

\subsection*{Magnetite synthesis by oxidizing greenrust in NaNO\textsubscript{2} solution}

The working matrix (400 mL) was prepared in a 500 mL \textit{Duran} glass bottle. After the solution was homogenized by stirring, 4 mL of FeCl\textsubscript{2} (0.5 mmol kg\textsuperscript{-1}) solution was added. The pH was adjusted to 8.0 by adding NaOH (1 mol kg\textsuperscript{-1}) solution drop-wise. After incubation for a week, the precipitate was filtered through 0.22 \textmu m membranes and separated into 3 portions mixed with different concentration of NaNO\textsubscript{2} solution (Table \ref{sample_prep}, entry 9, 10, 11). The resulting solutions were incubated inside the glovebox for another week. The precipitate was collected by filtration through 0.22 \textmu m membranes and dried in the low-humidity glovebox for 24 hours.

\subsection*{Magnetite synthesis by irradiating FeCl\textsubscript{2} solution}

The working matrix (400 mL) was prepared in a 500 mL \textit{Pyrex} glass beaker. After the solution was homogenized by stirring, 4 mL of FeCl\textsubscript{2} (0.5 mmol kg\textsuperscript{-1}) solution was added (Table \ref{sample_prep}, entry 1, 2, 3, three replicates.). The pH was adjusted to 8.0 by adding NaOH (1 mol kg\textsuperscript{-1}) solution drop-wise. The resulting solution was stirred with UV irradiation for 4 hours and then transferred to a \textit{Duran} glass bottle and sealed with a polypropylene cap. UV irradiation was performed using an \textit{Analytik Jena} UVP Pen-Ray mercury lamp (90-0004-01, 25 mA) placed above the solution. A control experiment has been run in parrallel without UV irradiation (table \ref{sample_prep}, entry 4). After one week of incubation, the precipitate was collected by filtration through 0.22 \textmu m membranes and dried in the low-humidity glovebox for 24 hours.

\begin{table}[ht]
\centering
\caption{Experimental conditions for magnetite synthesis.}
\begin{tabular}{llccccc}
\hline
\textbf{Entry} & \textbf{Sample No.} & 
\begin{tabular}{@{}c@{}} \textbf{FeCl\textsubscript{2}} \\ (mmol kg\textsuperscript{-1}) \end{tabular} & 
\begin{tabular}{@{}c@{}} \textbf{FeCl\textsubscript{3}} \\ (mmol kg\textsuperscript{-1}) \end{tabular} & 
\begin{tabular}{@{}c@{}} \textbf{NaNO\textsubscript{2}} \\ (mmol kg\textsuperscript{-1}) \end{tabular} &
\begin{tabular}{@{}c@{}} \textbf{Irradiation time} \\ (h) \end{tabular} &
\begin{tabular}{@{}c@{}} \textbf{Incubation time} \\ (d) \end{tabular} \\
\hline
1 & UV1    & 5 & 0 & 0 & 4 & 7\\
2 & UV2    & 5 & 0 & 0 & 4 & 7\\
3 & UV3    & 5 & 0 & 0 & 4 & 7 \\
4 & NUV4    & 5 & 0 & 0 & 0 & 7 \\
5 & NUV5   & 4.5 & 0.5 & 0 & 0 & 7\\
6 & NUV6   & 2.5 & 2.5 & 0 & 0 & 7\\
7 & NUV7   & 0.5 & 4.5 & 0 & 0 & 7\\
8 & NUV8   & 4.5 & 0.5 & 0 & 0 & 7\\
9 & NUV9\_1   & 5 & 0 & 0 & 0 & 7\\
10 & NUV9\_3   & 5 & 0 & 0.02 & 0 & 7\\
11 & NUV9\_4   & 5 & 0 & 0.1 & 0 & 7\\
\hline
\end{tabular}
\label{sample_prep}
\end{table}


\subsection*{Rock Magnetic Measurements}

Mass-specific magnetic susceptibility was measured at 976~Hz using an Agico Multifunction Kappabridge susceptibility meter (model MFK1-FA), with an applied field strength of 200~A/m. To minimize measurement uncertainty, each sample was measured five times, and the average susceptibility values were used. Magnetic hysteresis loops, backfield demagnetization curves and first-order reversal curves were measured using a Lakeshore PMC MicroMag 2900 alternating gradient field magnetometer (AGM). Hysteresis loops were acquired in discrete mode with a maximum applied field of 1~T and an increment of 2~mT. Hysteresis parameters (saturation magnetization $M_s$, remanent magnetization $M_{rs}$, and coercive force $B_c$) were determined after correcting for the high-field slope ($>700$~mT) to remove diamagnetic and paramagnetic contributions. The coercivity of remanence $B_{cr}$ was obtained by progressively demagnetizing a saturated sample in a reverse field down to $-1$~T. FORC diagrams were acquired at maximum applied fields of 1~T with 1.02~mT field increments and processed using FORCinel software (version 3.08) with smoothing parameters summarized in Table~S3 in the SI) \cite{Harrison2008, Egli2013}. Before processing, the lower branch of the hysteresis loop was subtracted from each FORC to generate a set of “difference FORCs.” Although this subtraction does not alter the FORC distribution, it significantly reduces common processing artifacts and removes sigmoidal contributions near the remanence diagonal, which cannot be adequately captured by the second-order polynomial function used to smooth the FORC surface \cite{Egli2013}. 

\subsection*{Electron microscopy and energy-dispersive spectroscopy}

Scanning electron microscopy (SEM) images were acquired using a \textit{FEI} Quanta 650 FEG SEM in high-vacuum mode. Backscattered electron (BSE) SEM images were acquired at accelerating voltages of 5--10~kV, while energy-dispersive X-ray spectroscopy (EDXS) elemental mapping was conducted at 5--15~kV. High-resolution BSE-SEM images (1~nm/pixel) were captured using AZtec software (version 6.2). For transmission electron microscopy (TEM) measurements, the powder was dispersed onto a holey carbon film supported on a 200-mesh copper grid from EMResolutions by suspending the powder in water, pipetting the suspension onto the grid and allowing the water to evaporate. The sample was imaged using Scanning TEM with an aberration corrected Thermo Fisher Scientific Spectra 300, operated with an accelerating voltage of 300 kV, in the Department of Materials Science and Metallurgy at the University of Cambridge. STEM images and EDS datasets were acquired simultaneously, using a beam current of approximately 100 pA and a convergence angle of 23 mrad, with a dwell time of 1 ms. EDS datasets were collected with the Super-X detector, consisting of four silicon drift detectors, and STEM images were collected with the Panther segmented STEM detector system, with a HAADF inner angle of 50 mrad. STEM and EDS images were analyzed using Hyperspy \cite{hyperspy_v2_3_0}. 

\subsection*{Scanning electron diffraction}

Scanning Electron Diffraction (SED) datasets were acquired also with the Thermo Fisher Scientific Spectra 300 and accelerating voltage as in the scanning TEM, with a 0.2 mrad convergence angle, a probe current of approximately 30 pA and a dwell time of 1 ms. SED patterns were recorded on a Quantum Detectors MerlinEM direct electron detector.

SED datasets were analysed using pxyem \cite{pyxem_v0_20_0}. The reciprocal space lengths in the diffraction patterns were calibrated based on patterns collected under the same conditions from AuPt nanoparticles. After centering the direct beam, a potential magnetite particle was identified based on its morphology, and the composition determined by EDS. Kinematical diffraction patterns were simulated using the Single Crystal 5 software, and compared with the summed diffraction patterns from the magnetite particle.

\subsection*{X-ray diffraction}

Powder X-ray diffraction (XRD) spectra were acquired using a \textit{Malvern Panalytical} Empyrean instrument with a Co K\textalpha{} source (1.78901 \AA) operated at 45.0 kV and 40.0 mA, scanned between 2 to 80 degrees 2\textTheta{} at a step size of 0.026 degrees and a total scan time of approximately 11 minutes. 

\subsection*{Micromagnetic modelling}

Micromagnetic simulations were performed to quantify how a CISS-active, spin-polarized molecular layer can bias the domain-state energetics of magnetite particles. In micromagnetics, the magnetization is represented by a continuous unit-vector field $\mathbf{m}(\mathbf{r})$, and quasi-static configurations are obtained by minimizing the total magnetic free energy, including exchange, magnetostatic (demagnetizing), magnetocrystalline anisotropy, and Zeeman terms \cite{Brown1963Micromagnetics}. Calculations were carried out using MERRILL, an open-source finite-element micromagnetic software suitable for natural magnetic minerals \cite{OConbhui2018MERRILL}. Room-temperature magnetite parameters were fixed to experimentally constrained values: exchange stiffness $A_{\mathrm{mag}} = 1.33\times10^{-11}$~J\,m$^{-1}$, magnetocrystalline anisotropy $K_{1} = -1.24\times10^{4}$~J\,m$^{-3}$, and spontaneous magnetization $M_{s} = 480$~kA\,m$^{-1}$ \cite{OConbhui2018MERRILL}.

\paragraph{Effective interfacial coupling to a chiral molecular layer} To incorporate the magnetic influence of a chiral molecular layer generated via chiral-induced spin selectivity (CISS), we adopt a deliberately coarse-grained interfacial description. The molecular layer is not treated as a magnetic medium with its own internal exchange or micromagnetic dynamics; instead, its net effect is represented as an interfacial energetic bias that favors one relative alignment between the magnetite surface magnetization and an effective molecular polarization axis $\mathbf{m}_{\mathrm{mol}}$. This formulation does not assume that the underlying microscopic mechanism is conservative or specify a particular non-equilibrium torque pathway; rather, it tests whether experimentally and electronically motivated interfacial energy scales are sufficient to reshape the particle’s quasi-static free-energy landscape and thereby modify domain-state stability and reversal pathways.

At the microscopic level, the interaction between an interfacial spin polarization associated with a CISS-active molecule and a surface spin may be represented by a Heisenberg-like exchange interaction,

\begin{equation}
H_{\mathrm{ex}} = -2J\,\mathbf{S}_{i}\cdot\mathbf{S}_{j},
\end{equation}

\noindent so that the energetic preference between antiparallel (AP) and parallel (P) alignment can be expressed as a per-interaction energy difference

\begin{equation}
\Delta E_{\mathrm{pair}} \equiv E_{\mathrm{AP}} - E_{\mathrm{P}} = 4JS^{2}.
\end{equation}

For an surface density $n_s$ of active interfacial interactions (molecules or spin-polarized carriers) on the magnetite surface, the corresponding AP--P energy difference per unit area is $n_s\,\Delta E_{\mathrm{pair}}$. In the micromagnetic implementation, we apply this bias by exchange-coupling the outermost magnetite elements to a fixed $\mathbf{m}_{\mathrm{mol}}$ across an effective interaction thickness $d$, taken as the thickness of the interfacial finite elements normal to the surface. In the continuum model, the interfacial bias can be represented by an exchange-like surface energy density of the form

\begin{equation}
E_{\mathrm{int}} \simeq \frac{2A_{\mathrm{exc}}^{\mathrm{IL}}}{d}\left(1-\mathbf{m}\cdot\mathbf{m}_{\mathrm{mol}}\right),
\end{equation}

\noindent which yields an antiparallel--parallel areal energy difference

\begin{equation}
\Delta E_{int} = \frac{4A_{\mathrm{exc}}^{\mathrm{IL}}}{d}.
\end{equation}

Equating $\Delta E_{int}$ to the microscopic areal preference $n_s\,\Delta E_{\mathrm{pair}}$ gives the mapping between experimentally constrained parameters and the micromagnetic coupling parameter:

\begin{equation}
\label{eq:Aexc_final}
A_{\mathrm{exc}}^{\mathrm{IL}}=\frac{n_s\,\Delta E_{\mathrm{pair}}\,d}{4}.
\end{equation}

Here, $\Delta E_{\mathrm{pair}}$ is the per-interaction energetic preference. Atomic force microscopy measurements have shown that such interaction can be as large as $\sim$150~meV. In our simulations, however, we assume a more conservative range of interactions from $\approx$0.01 to 100 meV to evaluate the magnetization changes in magnetite particles caused by the spin polarized layer. For a molecular density $n_s = 2 \times 10^{18}$ molecules per m\textsuperscript{2} \cite{ben2017magnetization, Ray1999} and an intermediate layer thickness $d = 1.0$~nm (i.e., spacing between magnetite and the spin polarized molecular layer in the micromagnetic mesh), $A_{exc}^{IL}$ varies from $8 \times 10^{-17}$ J/m to $8 \times 10^{-12}$ J/m, which corresponds to $\sim$~10\textsuperscript{-4} to  60\% of magnetite's internal exchange stiffness $A_{mag}$. 

In our simulations, the cumulative particle-level bias follows automatically by summing the interfacial contribution over the discretized surface. For comparison with experiments, we also define an equivalent bias field $B_{eff}$ as the uniform applied field that reproduces the same change in a chosen magnetic observable (e.g., remanent moment or switching threshold) as the interfacial coupling. For the particle sizes explored here (50--150~nm) and for the adopted $n_s$ and $d$, values of $\Delta E_{pair} \sim 0.1$--10~meV correspond to $B_{eff}$ of order $0.1$--$30$~mT, comparable to mT-scale biases inferred in adsorption-driven magnetic experiments \cite{ozturk2023central,Dor2013ChiralMemory}.

\paragraph{Scope and limitations of the model} The parametrization in Eq.~\eqref{eq:Aexc_final} is intentionally first-order. First, the microscopic complexity of CISS (spin--orbit coupling, charge redistribution/transport, electronic structure, and chemical bonding) is compressed into two effective quantities, $\Delta E_{\mathrm{pair}}$ and $n_s$, whose product sets the areal energetic bias. Second, the quasi-static energy-minimization approach isolates how an interfacial bias reshapes the free-energy landscape and the accessibility of metastable domain states, but does not, by itself, resolve non-equilibrium spin transport or explicit time-dependent switching kinetics. Third, $\mathbf{m}_{\mathrm{mol}}$ is treated as a fixed polarization axis and the coupling is applied uniformly over the surface, neglecting molecular-scale spatial heterogeneity. Despite these simplifications, the framework provides a transparent and physically grounded route to embed experimentally constrained interfacial energy scales into micromagnetic simulations of realistically sized magnetite grains.

\newpage

\end{document}